\pdfoutput=1
\documentclass[english,aps,prl,showpacs,twocolumn,superscriptaddress]{revtex4-1}
\usepackage{graphicx}
\usepackage{color}
\usepackage{calc}
\usepackage{calculator}
\usepackage{enumerate}
\usepackage{amsmath}
\usepackage{amssymb}
\usepackage{stmaryrd}
\usepackage{multirow}
\usepackage{tikz}
\usepackage{float}
\usepackage{breqn}
\usepackage{booktabs}
\usepackage{hyperref}
\usepackage{url}
\usepackage{soul}
\usepackage[caption=false]{subfig}
\begin{document}

\title{Disorder driven phase transitions in weak AIII topological insulators}

\author{Jahan Claes}
\author{Taylor L. Hughes}
\affiliation{%
 Department of Physics and Institute for Condensed Matter Theory$,$\\
 University of Illinois at Urbana-Champaign$,$ IL 61801$,$ USA
}%

\begin{abstract}
    The tenfold classification of topological phases enumerates all strong topological phases for both clean and disordered systems. These strong topological phases are connected to the existence of robust edge states. However, in addition to the strong topological phases in the tenfold classification, there exist weak topological phases whose properties under disorder are less well understood. It is unknown if the weak topological indices can be generalized for arbitrary disorder, and the physical signatures of these indices is not known. In this paper, we study disordered models of the two dimensional weak AIII insulator. We demonstrate that the weak invariants can be defined at arbitrary disorder, and that these invariants are connected to the presence or absence of bound charge at dislocation sites.
\end{abstract}

\maketitle
\section{Introduction}

The tenfold periodic table of topological phases enumerates all possible strong topological phases that are protected by internal symmetries and robust against disorder\cite{schnyder2008classification,kitaev2009periodic,qi2008,ryu2010topological,chiu2016classification}. The ten symmetry classes are distinguished by the presence or absence of time-reveral ($\hat{T}$), charge-conjugation ($\hat C$), and chiral ($\hat S$) symmetry. Within a given symmetry class, the insulating phases of Hamiltonians are categorized by a \emph{topological invariant} of that symmetry class, i.e., a quantized index that can be calculated from the many-body ground state of the Hamiltonian. Strong topological phases exhibit protected gapless edge modes that cannot be gapped by symmetry-preserving surface perturbations and are inherently robust to symmetry-preserving disorder\cite{bernevig2013topological,hasan2010colloquium,qi2011topological}.

Besides the strong topological phases enumerated in the tenfold periodic table, there also exist \emph{weak} topological phases\cite{fu2007topological,roy2009topological,moore2007topological,kitaev2009periodic}. These phases nominally require lattice translation symmetry for their protection and can be viewed as ``stacks" of lower dimensional topological insulators.  The corresponding weak topological invariants are formed by averaging lower-dimensional strong topological invariants across the stack. For example, in the chiral-symmetric AIII class, the tenfold periodic table shows there exists a strong $\mathbb{Z}$-valued topological index for 1D systems, the winding number $\nu$ (see below for further discussion of $\nu$), and no strong topological index for 2D systems. However, a 2D system formed from a stack of $N_y$ 1D chains with $\nu=1$ will have a weak topological index $\nu_x$ given by the average winding, $\nu_x=\nu_{total}/N_y=1$. Weak topological phases also exhibit gapless edge modes for particular edge terminations that are compatible with the stacking direction, and additionally bind gapless states at crystalline defects such as dislocations\cite{ran2009one,teo2010topological,teo2017topological}.

Explicit expressions for calculating the strong topological invariant are generally given in terms of integrals over the Brillouin zone, and thus seemingly require translational invariance. However, the use of momentum space is just a basis choice, and formulas for the strong topological invariants in most symmetry classes have subsequently been generalized to real-space formulas that apply even to disordered systems \cite{prodan2010,loring2011,mondragon2014topological,prodan2016non,prodan2017computational,schulz2016topological,prodan2016bulk,prodan2013non}. These real-space topological invariants recover the usual invariants when applied to a translationally invariant system. Furthermore, they are stable against weak disorder, are quantized even in the presence of strong disorder, and many of them can change values only if delocalized states exist at the Fermi level.

Ref. \onlinecite{prodan2016bulk} also established real-space formulae for \emph{weak} topological invariants, and demonstrated that the corresponding invariants were quantized and stable for weak disorder as long as the spectral gap remains open. Recent mathematical work using KK-theory\cite{prodan2016generalized} and semifinite index theory\cite{bourne2018application} have also suggested formulae to calculate real-space weak topological invariants and demonstrated their quantization and stability under weak disorder. However, in all cases the properties of these weak topological invariants under strong disorder are not known.

In this paper, we explicitly study some classes of weak topological insulators in weak and strong disorder regimes. We define our weak topological invariants by directly apply the corresponding formulae for strong invariants in lower dimensions, as was suggested in \cite{prodan2016bulk}. Our focus will be on 2D insulators in class AIII where the previously-known\cite{mondragon2014topological} real-space formula for the 1D winding number $\nu$ allows us to compute the average 2D winding $\nu_x\equiv \nu_{\text{total}}/N_y$, by treating the 2D system as a 1D system with a width $N_y.$ In this approach, the stability of these indices for weak disorder immediately follows from results on the corresponding strong invariants. However, the behavior at strong disorder is still not known. The theorems in Refs. \onlinecite{prodan2016non,prodan2017computational,schulz2016topological,prodan2016bulk} establish that the strong invariants are quantized to integer values, but this only proves that a weak invariant like $\nu_x$ is quantized in units of $\frac{1}{N_y}$ at strong disorder, not that $\nu_x$ is itself an integer. Thus at strong disorder the quantization could vanish in the thermodynamic limit.

Using our real-space invariants, we study the phase diagrams of two models of the chiral-symmetric AIII class in two dimensions. In two dimensions, this class has two weak topological invariants, $\nu_x$ and $\nu_y$, and no strong topological invariants. We confirm through computations of the weak topological invariants, and numerical transfer matrix methods, that the weak topological indices are robust against weak disorder. We also demonstrate numerically that the weak invariant remains quantized (in integer units) even at strong disorder when the Fermi level lies in a region of localized eigenstates, a result that has no known analytic proof. We show that generically, the disorder-driven phase transition in 2D is strikingly different from the 1D strong-topological version\cite{mondragon2014topological,song2014aiii}. Rather than a sharp phase boundary separating regions of different winding, there is a continuous transition region of delocalized states where the averaged winding varies smoothly between different values; this is reminiscent of a Dirac semi-metal phase separating the weak topological phase and trivial phase that could appear in the clean limit. Finally, we demonstrate a connection between the real-space weak topological invariant and physical observables, namely the electronic polarization and the bound charge at a dislocation.

\section{Background}

To begin, we review the topological properties the AIII class, for both clean and disordered systems. We pay particular attention to the AIII class in 1D and 2D, which will be relevant for our work.

The AIII class is defined to be the set of all Hamiltonians that anticommute with an on-site unitary operator $\hat{S}$ satisfying $\hat{S}^2=\mathbf{1}$, called the chiral symmetry operator. This implies that
\begin{equation}
    \hat{S}\hat{H}\hat{S}=-\hat{H}.
    \label{chiralSymm}
\end{equation}
An immediate consequence of this equation is that the spectrum of $\hat{H}$ is symmetric about $E=0$, since if $|\psi_n\rangle$ is an eigenstate of $\hat{H}$ with energy $E_n$, then $\hat{S}|\psi_n\rangle$ is an eigenstate of $\hat{H}$ with energy $-E_n$. The Fermi energy $E_F$ is assumed to be $0$, which is compatible with the chiral symmetry and equivalent to assuming half filling due to the symmetry of the spectrum.

A simple example of a chiral symmetry operator in tight-binding models is the sublattice operator. For a given bipartite tight-binding lattice, we may divide the lattice into two sublattices A and B. In terms of this division, we define a sublattice operator that flips the sign of a single sublattice
$$
\hat{S} =\sum_i \left( c_{i,A}^\dagger c_{i,A}-c_{i,B}^\dagger c_{i,B}\right).
$$
Clearly $\hat{S}^2=\mathbf{1}$. A simple calculation shows that $\hat{S}\hat{H}\hat{S}=-\hat{H}$ whenever $\hat{H}$ does not include hoppings within a single sublattice, such as $c_{i,A}^\dagger c_{j,A}$. The examples of such tight-binding Hamiltonians that we will study are illustrated in Fig. \ref{models}. Each of these lattice models is bipartite, and their chiral  symmetry operator will be the corresponding sublattice operator.

For translationally invariant systems in 1D we can define the topological invariant in the AIII class as follows. First, Eq. \ref{chiralSymm} can be expressed in $k$-space for the Bloch Hamiltonian as
\begin{equation}
    \hat{S}\hat{H}(\vec k)\hat{S}=-\hat{H}(\vec k).
    \label{chiralSymmTrans}
\end{equation}
Since $\hat{S}^2=\mathbf{1}$, the eigenvalues of $\hat{S}$ are $\pm 1$. If $|u_\pm(\vec{k})\rangle$ satisfies $\hat{S}|u_\pm(\vec{k})\rangle=\pm |u_\pm(\vec{k})\rangle$, then Eq. \ref{chiralSymm} implies $\hat{S}\hat{H}(\vec{k})|u_\pm(\vec{k})\rangle=\mp \hat{H}(\vec{k})|u_\pm(\vec{k})\rangle$. In other words, $H(\vec{k})$ takes $+1$ eigenstates of $\hat{S}$ to $-1$ eigenstates of $\hat{S}$, and thus takes an off-block-diagonal form when expressed in the basis of $\hat{S}$ eigenstates:
\begin{equation}
    \hat{H}(k)=\left(\begin{array}{cc}0& \hat{q}(k)\\\hat{q}^\dagger(k)&0\end{array}\right)
    \label{BlockOffDiagonal}
\end{equation}
for some operator $\hat{q}(k)$. The 1D winding number invariant takes the form:
\begin{equation}
    \nu = -\frac{i}{2\pi}\int dk\ \text{Tr}\left[q^{-1}\partial_k q\right].
    \label{1dWindingTranslInv}
\end{equation}

Ref. \onlinecite{mondragon2014topological} introduced a covariant real-space generalization of $\nu$ to AIII systems without translational symmetry. The key ingredient in defining the real-space topological invariant is a ``flat band Hamiltonian" defined by
\begin{equation}
    \hat{Q}\equiv 1-2\hat{P},
\end{equation}
where $\hat P$ is the projector onto the occupied states. By the same reasoning that led to Eq. \ref{BlockOffDiagonal}, when written in a basis of eigenstates of $\hat{S}$, this flat-band Hamiltonian again takes an off-block-diagonal form
\begin{equation}
    \hat{Q}=\left(\begin{array}{cc}0& \hat{Q}_{+-}\\\hat{Q}_{-+}&0\end{array}\right),
\end{equation}
where $\hat{Q}_{-+}=\hat{Q}_{+-}^\dagger$. In terms of $\hat{Q}$, the real-space generalization of Eq. \ref{1dWindingTranslInv} is given by
\begin{equation}
    \nu=-\frac{1}{N_x}\text{Tr}\left\{\hat{Q}_{-+}[\hat{X},\hat{Q}_{+-}]\right\}
    \label{1dWindingReal}
\end{equation}\noindent where $N_x$ is the number of unit cells in the $x$-direction, and $\hat{X}$ is the position operator.

This real-space $\nu$ reduces to the original winding number for translationally invariant systems. It has been demonstrated both numerically\cite{mondragon2014topological,song2014aiii} and analytically\cite{prodan2016non} that $\nu$ is quantized to integer values even in the presence of strong disorder, provided only localized eigenstates exist at the Fermi level $E_F=0$. In addition, Refs. \onlinecite{mondragon2014topological,song2014aiii} demonstrated that the transitions between different values of $\nu$ are sharp. From these properties we would expect the phase diagram in disorder-space to consist of regions of quantized $\nu$ separated by codimension-1 phase boundaries, and indeed, this is exactly what one finds in 1D\cite{mondragon2014topological,song2014aiii}.

In the 2D AIII class, no strong topological invariants exist. However, there are topological classes in 2D determined by the weak topological indices $\nu_x$ and $\nu_y$. These weak 2D topological indices can be defined in terms of the strong 1D topological index. We define $\nu_x$ by treating our two-dimensional system as a one-dimensional system that is infinite in the $x$-direction and with width $N_y$ in the $y$-direction. Then $\nu_x$ is defined as the 1D winding per unit width, $\nu_x=\nu/N_y$. By dividing by $N_y$, we ensure $\nu_x$ is well-defined in the limit $N_y\rightarrow \infty$. The weak invariant $\nu_y$ is defined similarly, by treating the system as a one-dimensional system infinite in the $y$-direction and with width $N_x$ in the $x$-direction, and dividing the resulting winding by $N_x$.
For translationally invariant gapped systems, $\nu_x$ and $\nu_y$ are still integers, despite the fact that the original integer-valued invariant $\nu$ has been divided by $N_x$ or $N_y$. The indices $\nu_x$ and $\nu_y$ individually depend on the choice of basis vectors $\hat{x}$ and $\hat{y}$ of the lattice, but the vector quantity
\begin{equation}
    \vec\nu = \nu_x \hat{x}+\nu_y\hat{y}
    \label{nuvector}
\end{equation}
is independent of the basis vectors\cite{fu2007topological}.

With these definitions the resulting real-space formulas for the weak invariants are then
\begin{equation}
\begin{aligned}
    &\nu_x=-\frac{1}{N_xN_y}\text{Tr}\left\{\hat{Q}_{-+}[\hat{X},\hat{Q}_{+-}]\right\},\\
    &\nu_y=-\frac{1}{N_xN_y}\text{Tr}\left\{\hat{Q}_{-+}[\hat{Y},\hat{Q}_{+-}]\right\}.
\end{aligned}
    \label{weakRealSpaceInvariants}
\end{equation}
Note that while $\nu$ must be integer valued, $\nu_x$ and $\nu_y$ need not be, even if the Fermi level is in a region of localized states. Rather, since $\nu_x$ and $\nu_y$ are obtained from $\nu$ by dividing by $N_y$ and $N_x$, respectively, $\nu_x$ is only guaranteed by Ref. \onlinecite{prodan2016non} to be quantized to integer multiples of $\frac{1}{N_y}$, and similar for $\nu_y$. In the thermodynamic limit, then, the $\nu_\alpha$ can in principle take on any value in $\mathbb{R}$. One of the important results of this work will be demonstrating that when the Fermi level is in a region of localized eigenstates, the $\nu_\alpha$ retain their quantization and are nonetheless still integers even as one approaches the thermodynamic limit.

\section{2D AIII Models}

\begin{figure}
    \centering
    \subfloat[SSH]{
        \includegraphics[width=.9\columnwidth]{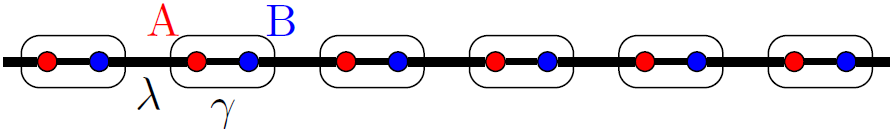}
        \label{SSHModel}
    }
    
    \subfloat[$\hat H_1$]{
        \includegraphics[width=.43\columnwidth]{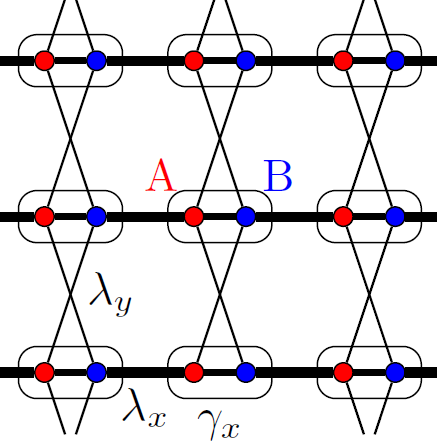}
        \label{H1Model}
    }
    \subfloat[$\hat H_2$]{
        \includegraphics[width=.43\columnwidth]{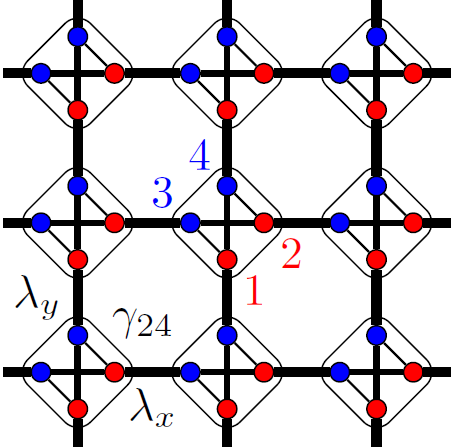}
        \label{H2Model}
    }
    
    \caption{(a) The 1D chiral symmetric SSH chain, the basic building block for our models. (b) Model $\hat H_1$ consists of stacked SSH chains, with all nearest-neighbor hoppings compatible with chiral symmetry. (c) Model $\hat H_2$ consists of crossed SSH chains, with all intracell hoppings compatible with chiral symmetry.}
    \label{models}
\end{figure}

The simplest way to form two-dimensional models with nontrivial weak indices is to stack 1D models with nontrivial $\nu$. Our basic building block will be the chiral symmetric 1D Su-Schrieffer-Heeger (SSH) chain\cite{su1979solitons}, the simplest example of an AIII topological insulator in 1D (Fig. \ref{SSHModel}). The SSH chain has $\nu=1$ when $|\lambda|>|\gamma|$, and $\nu=0$ otherwise. To create our 2D models, we first stack SSH chains, and then add some local inter-chain couplings that are allowed by the chiral (sublattice) symmetry.

The two 2D models we consider are illustrated in Figs. \ref{H1Model} and \ref{H2Model}. The two sublattices A and B are colored red and blue, respectively; note that there is no coupling between sites on the same sublattice, as required by chiral symmetry. Our first model, shown in Fig. \ref{H1Model}, is a two-band model given by a vertical stack of SSH chains in which nearest-neighbor unit cells are coupled in the vertical direction. Our second model, shown in Fig. \ref{H2Model}, is a four-band model given by two crossed stacks of SSH chains, with intracell couplings between the horizontal and vertical chains. While the pattern of hoppings may appear artificial, such hopping patterns can be realized experimentally in metamaterials, such as topolectric circuits \cite{lee2018topolectrical}. In addition, for solid-state realizations the A and B sites don't need correspond to physical lattice sites; they may refer to any degrees of freedom located in a unit cell\footnote{One simple way to realize such a hopping pattern in a solid-state context is to consider a spinless superconductor with time-reversal symmetry $\hat{\mathcal{T}}=\hat{K}$. Then the product of time-reversal and particle-hole symmetry gives a chiral symmetry\cite{chiu2016classification}. In this case, our $\hat{H}_1$ corresponds to a superconductor with one lattice site per unit cell, the degrees of freedom A and B correspond to Bogoliubov quasiparticle operators $(c_i\pm c_i^\dagger)/\sqrt{2}$, and coupling between them is forbidden simply by $\hat{\mathcal{T}}$ symmetry. Note that adding $\hat{\mathcal{T}}$ symmetry technically puts us in the BDI rather than AIII class, but the topological invariants of these two classes are identical\cite{chiu2016classification,song2014aiii}.}.

Written explicitly, the Hamiltonians are given by
\begin{multline}
    H_1=\sum_{\vec{r}}\left\{ \gamma_x^{\vec{r}} c_{\vec{r},A}^\dagger c_{\vec{r},B}+\lambda_y^{\vec{r},A} c_{\vec{r}+\hat{y},B}^\dagger c_{\vec{r},A} \right.\\\left.
    +\lambda_y^{\vec{r},B} c_{\vec{r}+\hat{y},A}^\dagger c_{\vec{r},B}
     +\lambda_x^{\vec{r}}c^\dagger_{\vec{r}+\hat{x},A}c_{\vec{r},B}+\text{h.c.}\right\},
\end{multline}
\begin{multline}
    H_2=\sum_{\vec{r}}\left\{ \lambda_x^{\vec{r}}c_{\vec{r}+\hat{x},3}^\dagger c_{\vec{r},2}+ \lambda_y^{\vec{r}} c_{\vec{r}+\hat{y},1}^\dagger c_{\vec{r},4}+ \gamma_{24}^{\vec{r}}c_{\vec{r},4}^\dagger c_{\vec{r},2}+ \right.\\\left. \gamma_{23}^{\vec{r}}c_{\vec{r},3}^\dagger c_{\vec{r},2}+ \gamma_{14}^{\vec{r}}c_{\vec{r},4}^\dagger c_{\vec{r},1}+ \gamma_{13}^{\vec{r}}c_{\vec{r},3}^\dagger c_{\vec{r},1}+\text{h.c.} \right\}.
\end{multline}
In both models, the parameters $\lambda_\alpha^{\vec{r}}$ and $\gamma_\alpha^{\vec r}$ represent position dependent hoppings. We study disordered versions of translationally-invariant systems, so we choose hoppings via
\begin{equation}
    \lambda_\alpha^{\vec{r}} = \lambda_\alpha^0+W_\lambda\omega^{\vec{r}}_{\lambda\alpha}\qquad    \gamma_\alpha^{\vec{r}} = \gamma_\alpha^0+W_\gamma\omega^{\vec{r}}_{\gamma\alpha}.
\end{equation}
Here, $\lambda_\alpha^0$ and $\gamma_\alpha^0$ are the hopping strengths in the clean limit, the $\omega_{\gamma\alpha}^{\vec{r}}$ and $\omega_{\lambda\alpha}^{\vec{r}}$ are independent random variables uniformly distributed in $[-0.5,0.5]$, and $W_\gamma$ and $W_\lambda$ are measures of the intracell and intercell disorder strengths, respectively.

\begin{figure}
    \centering
    \subfloat[$\hat H_1$]{
        \includegraphics[width=.975\columnwidth]{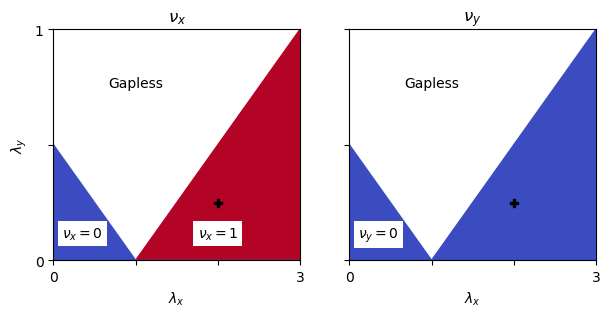}
    }
    
    \subfloat[$\hat H_2$]{
        \includegraphics[width=.975\columnwidth]{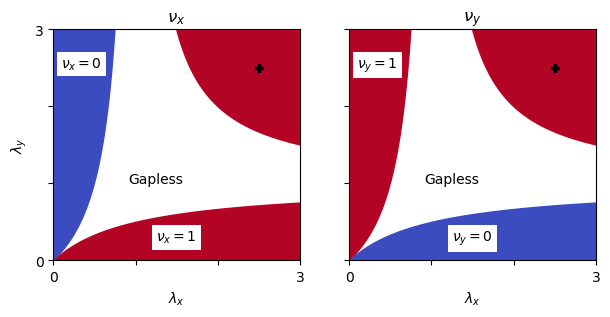}
    }
    \caption{(a) The $(\nu_x,\nu_y)$ phase diagram for the clean limit of $\hat H_1$. Here, we set all $\gamma_\alpha=1$. (b) The same phase diagram for $\hat H_2$. In both cases, the cross denotes the clean parameters $\lambda_\alpha^0$ we choose for our models.}
    \label{cleanPhase}
\end{figure}

In the clean limit, $\hat H_1$ is gapped provided either $|\gamma_x|+2|\lambda_y|<|\lambda_x|$ or $|\lambda_x|+2|\lambda_y|<|\gamma_x|$, and gapless otherwise. In the first gapped case $(\nu_x,\nu_y)=(1,0)$, while in the second gapped case $(\nu_x,\nu_y)=(0,0)$. Meanwhile, $\hat H_2$ is gapped provided $|\lambda_x\lambda_y|>|\gamma \lambda_x|+|\gamma \lambda_y|$, or $|\gamma\lambda_y|>|\gamma \lambda_x|+|\lambda_x\lambda_y|$, or $|\gamma\lambda_x|>|\gamma \lambda_y|+|\lambda_x\lambda_y|$, and gapless otherwise. In the first gapped case, $(\nu_x,\nu_y)=(1,1)$, while in the second and third gapped cases $(\nu_x,\nu_y)=(0,0)$. The phase diagrams for each model in the clean limit are shown in Fig. \ref{cleanPhase}. For our disordered systems we choose the initial parameters $(\lambda_x^0,\lambda_y^0,\gamma_x^0)=(2,0.25,0.5)$ for $\hat H_1$ and $(\lambda_x^0,\lambda_y^0,\gamma^0)=(2.5,2.5,1)$ for $\hat H_2$. The parameters represent arbitrary points in the topological phase, and can be changed without affecting our conclusions. These points in the phase diagram are indicated by black crosses in Fig. \ref{cleanPhase}, where it is apparent that we start with $(\nu_x,\nu_y)=(1,0)$ and $(1,1)$ for $\hat H_1$ and $\hat H_2$, respectively. Note that systems with arbitrary $(\nu_x,\nu_y)$ can be formed from stacks of these two elementary models.

\section{Disordered Weak Topological Insulators}

\subsection{Behavior for finite-width systems}

\begin{figure}
    \centering
    \includegraphics[width=\columnwidth]{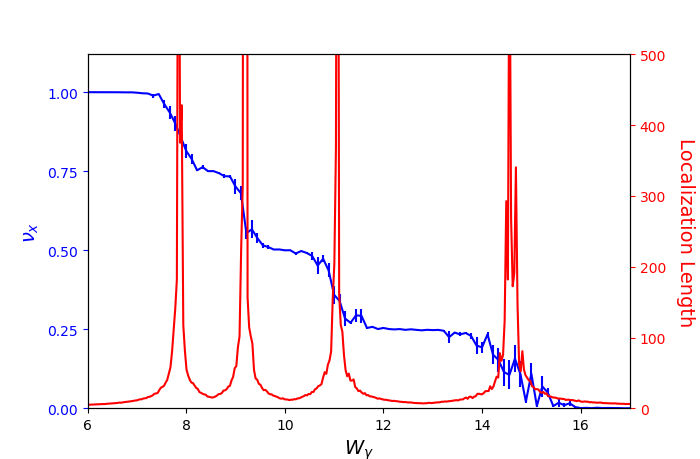}
    \caption{An example of the relationship between localization length and $\nu_x$ for model $\hat{H}_1$ with $N_y=4$. Here, we set $W_\lambda=0$ and tune $W_\gamma$. $\nu_x$ is plotted in blue, while the localization length $\Lambda_x$ of states at $E_F=0$ is overlaid in red. $\nu_x$ has been averaged over $\mathcal{N}_s=10$ disorder configurations. We see that $\nu_x$ is quantized in units of $\frac{1}{N_y}=.25$, and jumps at isolated points where the localization length goes to infinity. In the limit $N_y\rightarrow\infty$, this will translate to $\nu_x$ continually changing in \emph{regions} where the localization length is infinite, and being constant in regions where the localization length is finite (see Figs. \ref{resultsModel1},\ref{resultsModel2}).}
    \label{Example}
\end{figure}

In order to connect our results to Refs. \onlinecite{mondragon2014topological,song2014aiii}, we first explore the behavior of $\nu_x$ for disordered systems with finite $N_y$ in the limit $N_x\rightarrow\infty$. We can then treat our system as a thick 1D chain of width $N_y$. From the work in Refs. \onlinecite{mondragon2014topological,song2014aiii}, we expect that the 1D topological invariant $\nu$ given by Eq. \ref{1dWindingReal} is quantized to integer values, except at points in in the phase diagram where the localization length at $E_F=0$ diverges. We also know that as $W_\gamma\rightarrow\infty$, we should have $\nu=0$, since in this limit the intracell terms in the Hamiltonian dominate, and we thus expect a trivial insulator. This can also be seen directly from Eq. \ref{1dWindingReal}, since in this limit $\hat{Q}$ commutes with $\hat{X}$. We thus expect that as we increase $W_\gamma$ from 0, $\nu$ will decrease in some manner from $N_y$ to $0$, and any drops in $\nu$ should be accompanied by diverging localization length diverges at $E_F=0.$ In terms of our weak topological index $\nu_x$  we expect similar behavior. Namely, we expect $\nu_x=1$ at zero disorder, $\nu_x=0$ for $W_\gamma\rightarrow\infty$, and $\nu_x$ decreases as we increase $W_\gamma$.

We show the results of our numerical calculations in Fig. \ref{Example} for $\hat{H}_1$ with a width $N_y=4$. We see that $\nu_x$ drops in steps of size $\frac{1}{N_y}=.25$, and these drops occur precisely when the localization length diverges. This result shows that the transition between a (finite-width) weak topological insulator and a trivial insulator is generically split into $N_y$ separate transitions in the presence of disorder, and each transition is accompanied by a diverging localization length in the $x$-direction. 

We these results in mind, we now want to consider novel features that may emerge in the $N_y\rightarrow\infty$ limit. In this limit, we might expect that the $N_y$ isolated values of $W_\gamma$ at which the localization length diverges coalesce into an interval (or, in the case of higher-dimensional disorder configuration space, a region) having a diverging localization length throughout. In such an interval (or region), the value of $\nu_x$ can change continuously. However, if there are intervals or regions in disorder-space where the localization length remains finite as $N_y\rightarrow\infty$, $\nu_x$ cannot change in these regions. When considering the limit $N_y\rightarrow\infty$, we will be interested in the following questions:

\begin{itemize}
    \item Are there still sharp transitions between different values of $\nu_x$ as in Refs. \onlinecite{mondragon2014topological,song2014aiii}, or do points with diverging localization length coalesce into a region of disorder space, allowing $\nu_x$ to change continuously in this region?
    \item Do there still exist regions of disorder-space where the localization length remains finite, where $\nu_x$ plateaus to a well-defined value as $N_y\rightarrow\infty$?
    \item If there still exist regions of disorder-space with finite localization length, can $\nu_x$ plateau at any value, or does $\nu_x$ take on only integer values as in the clean limit?
\end{itemize}

\subsection{Disordered Phase Diagrams in the Thermodynamic Limit}
\begin{figure}
    \centering
    \includegraphics[width=\columnwidth]{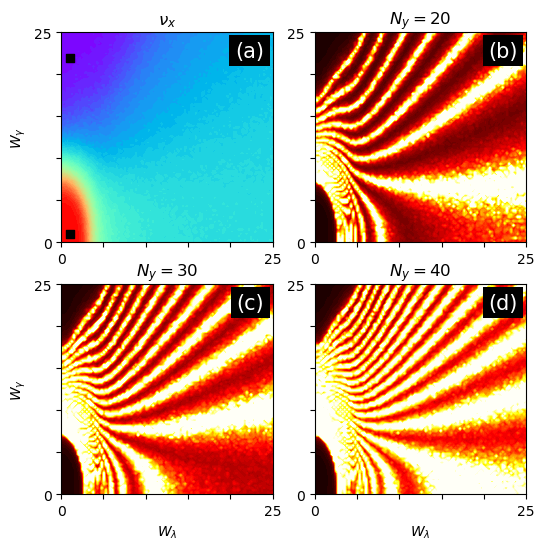}
    \caption{(a) $\nu_x$ as a function of disorder parameters $W_\gamma$ and $W_\lambda$ for $\hat H_1$. The results have been averaged over $\mathcal{N}_s=25$ disorder configurations. (b-d) the localization length $\Lambda_x$ as a function of $W_\gamma$ and $W_\lambda$, for increasing values of the width $N_y$. We see that as $N_y\rightarrow \infty$, we obtain islands of localized states separated by regions of delocalized states, and $\nu_x$ remains integer-valued on these localized islands.}
    \label{resultsModel1}
\end{figure}

\begin{figure}
    \centering
    \includegraphics[width=\columnwidth]{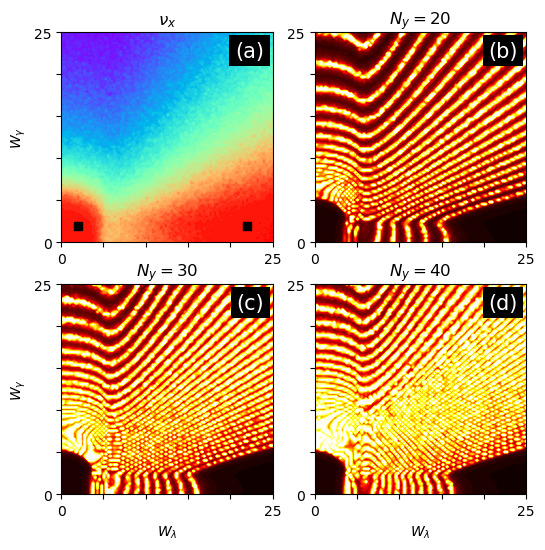}
    \caption{(a): $\nu_x$ as a function of disorder parameters $W_\gamma$ and $W_\lambda$ for $\hat H_2$. The results have been averaged over $\mathcal{N}_s=7$ disorder configurations. (b-d) the localization length $\Lambda_x$ as a function of $W_\gamma$ and $W_\lambda$, for increasing values of the width $N_y$. We again see that as $N_y\rightarrow \infty$, we obtain islands of localized states separated by regions of delocalized states, and $\nu_x$ remains integer-valued on these localized islands.}
    \label{resultsModel2}
\end{figure}

In Figs. \ref{resultsModel1}a and \ref{resultsModel2}a we plot phase diagrams  for models $\hat H_1$ and $\hat H_2$, respectively, as a function of two disorder parameters. Here, as in Fig. \ref{cleanPhase}, red (blue) represents $\nu_x=1 (0)$. For both $\hat H_1$ and $\hat H_2$, we find that $\nu_x$ is smooth as a function of disorder, and we do not see a sharp transition between values of $\nu_x$ that are quantized to integers. However, in both figures we find a region around zero disorder where $\nu_x$ remains quantized at 1 up to a finite value of disorder. In addition, in both cases we see ``islands" at strong disorder where $\nu_x$ regains quantization; in the case of $H_1$ an island having $\nu_x=0$ appears at strong $W_\gamma$, and in the case of $H_2$ an island having $\nu_x=1$ appears at strong $W_\lambda$. Finally, at no point do we see plateaus at non-integer values that survive the $N_y\rightarrow\infty$ limit.

Let us address possible concerns that the above features are just artifacts of finite-size effects. One might worry that the phase transition becomes sharp in the thermodynamic limit, or that the islands of quantized $\nu_x$ disappear. We can eliminate these possibilities by considering the localization length at $E_F=0$. Because the topological invariant can only change when there are delocalized states at $E_F=0$,  computing the localization length at $E_F=0$ can constrain the points in the phase diagram where $\nu_x$ is allowed to change.
To compute the localization length, we use a transfer-matrix method that numerically determines the localization length for samples which are finite in the $N_y$ direction, but arbitrarily large in $N_x$\cite{pichard1981finite}. The results are plotted in Figs. \ref{resultsModel1}(b-d) and \ref{resultsModel2}(b-d) for progressively larger values of $N_y$. In both cases, the behavior in the $N_y\rightarrow\infty$ limit is readily apparent. We observe that delocalized states coalesce to densely fill regions of the disorder configuration space, but are excluded from the islands where we claim that $\nu_x$ remains quantized. These localization calculations demonstrate that, even in the $N_y\rightarrow\infty$ limit, $\nu_x$ must remain quantized for weak disorder, as there are no delocalized states at weak disorder. It also demonstrates that no islands of stable, non-quantized values of $\nu_x$ appear in the phase diagram as $N_y\rightarrow\infty$, as the only regions we find without delocalized states are regions in which $\nu_x$ is integer-valued (i.e., $\nu_x\in\mathbb{Z}$ and is not just an integer multiplying $1/N_y$).

Let us summarize these results. Our real-space weak invariants remain quantized at weak disorder that does not close the bulk gap, as in Refs. \onlinecite{prodan2013non,prodan2016generalized,bourne2018application}. In our formulation, this is manifest because our formula for $\nu_x$ cannot change without delocalized states appearing at the $E_F=0$, and at weak disorder the spectral gap remains open and $\nu_x$ cannot change.
In addition, we have numerical evidence that the weak indices remain quantized even at strong disorder provided there exist no delocalized states at the Fermi level, something that analytic arguments have not yet established. We conjecture that this is generic, i.e., that $\nu_x$ and $\nu_y$ are quantized for \emph{any} homogeneous disorder that obeys chiral symmetry and does not result in delocalized states at the Fermi level. We also find that there is a contiguous region of delocalized states separating the weak topological phase from the trivial insulator phase, and as we will see below, the properties of this critical region are reminiscent of those expected for an intermediate Dirac semimetal phase that separates the weak from the trivial phase\cite{ramamurthy2015}. In this phase, the metallic character is robust in a region of the phase diagram, and is protected from disorder by the fractional value of $\nu_x$. 

The quantization of $\nu_x$ at strong disorder allows us to identify highly-disordered systems as having protected weak topological properties. Indeed, once we connect $\nu_x$ to real-space observables (see below), the quantization of $\nu_x$ at strong disorder implies that even strongly disordered systems can exhibit stable, quantized properties that are protected by real-space weak topology.

\subsection{Signatures of Disordered Weak Topological Insulators}

\subsubsection{Polarization and edge modes}

For a clean AIII system, if we introduce a clean boundary perpendicular to $\hat{x}$, the invariant $\nu_x$ predicts the number of robust zero-energy modes (per $N_y$) localized at the boundary. These robust zero-energy modes are chiral, satisfying $S|\psi_n\rangle=\pm|\psi_n\rangle$ with the same sign for all the modes on the boundary. This implies that these states cannot be moved away from zero energy by any chiral-symmetric perturbation, since the matrix elements between them necessarily vanish: $\langle\psi_m|\hat{H}_{\text{pert}}|\psi_n\rangle=\langle\psi_m|\hat{S}\hat{H}_{\text{pert}}\hat{S}|\psi_n\rangle=-\langle\psi_m|\hat{H}_{\text{pert}}|\psi_n\rangle=0.$ In addition to providing a spectroscopic signature of the weak topological phase these modes result in the manifestation of a boundary charge generated by a polarization given by $2p_x=e\nu_x \text{ mod } 2e$ where $e$ is the electric charge and we have assumed lattice constants are normalized to unity.

With the addition of weak disorder that does not close the bulk gap, $\nu_x$ still robustly predicts the number of edge modes, and thus can be associated with a bulk polarization. For stronger disorder, we cannot easily distinguish between the protected zero energy modes on the edges and the bulk zero energy modes. However, a signature of the zero energy modes remains. If we add a small chiral-symmetry (and inversion-symmetry) breaking term $\delta \hat{S}$ to the Hamiltonian, the charge density per unit length on an edge is given by $\pm e\nu_x/2$, where the sign depends on the sign of $\delta$. Remarkably, we find that this holds even for the case of fractional $\nu_x$, when delocalized states exist at zero energy, similar to the well-defined polarization and boundary charge in a gapless 2D Dirac semi-metal\cite{ramamurthy2015}. Therefore, the polarization of a sample with edges\footnote{Note that we calculate the polarization with open boundary conditions through the surface charge. Methods for computing the polarization with periodic boundary conditions such as those of Refs. \onlinecite{resta1998quantum,ortiz1994macroscopic} only give the dipole moment per length $N_x$ mod $e$, and cannot determine the dipole moment per unit volume.} still satisfies $2p_x=e\nu_x \text{ mod } 2e$.

\subsubsection{Bound charge at dislocations}

In clean systems, it is known that weak topological insulators have protected modes at edge (and in 3D, screw) dislocations\cite{chiu2016classification,teo2010topological,ran2009one}. In 2D AIII systems without disorder, a dislocation hosts a bound charge $q_b$ that obeys the relationship
\begin{equation}
    \frac{e}{2}\vec{\nu}\times \vec{b} = \pm q_b
    \label{boundChargeEqn}
\end{equation}
where $\vec{b}$ is the Burgers vector\cite{kittel1976introduction} of the dislocation, and $\vec\nu$ is as the weak invariant defined in Eq. \ref{nuvector}. For clean systems, this bound charge cannot be removed by (weak) disorder near the edge dislocation; thus, these modes are a possible signature of the weak topological phase even with bulk disorder.

\begin{figure}
    \centering
    \subfloat[$\hat H_1$ at the point $(W_\lambda,W_\gamma)=(1,1)$, where $\nu_x=1$.]{
        \includegraphics[width=.9\columnwidth]{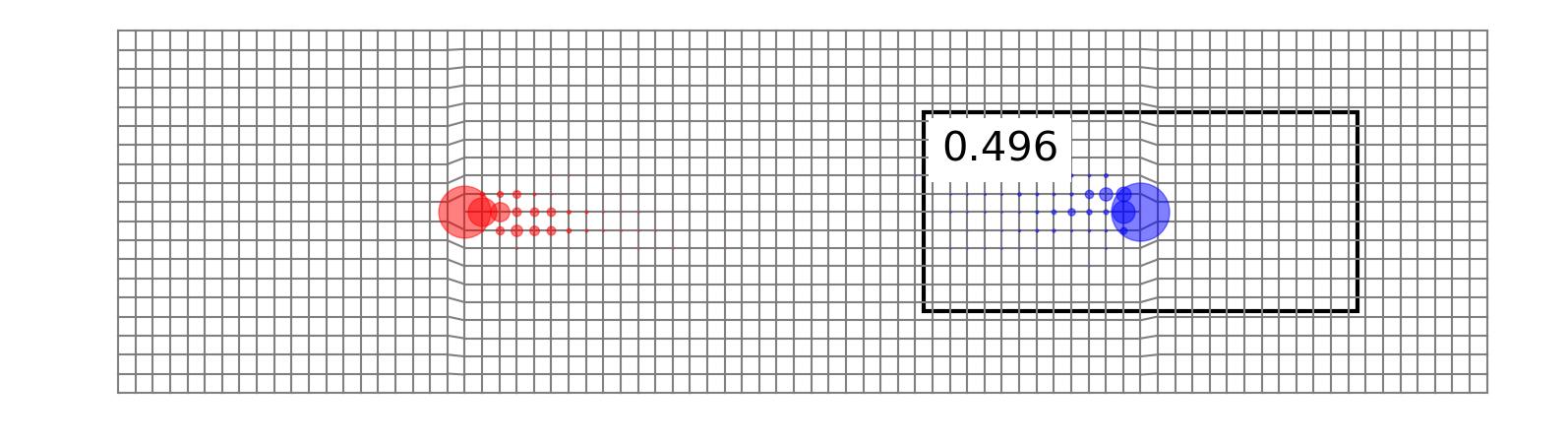}
    }

    \subfloat[$\hat H_1$ at the point $(W_\lambda,W_\gamma)=(1,22)$, where $\nu_x=0$.]{
        \includegraphics[width=.9\columnwidth]{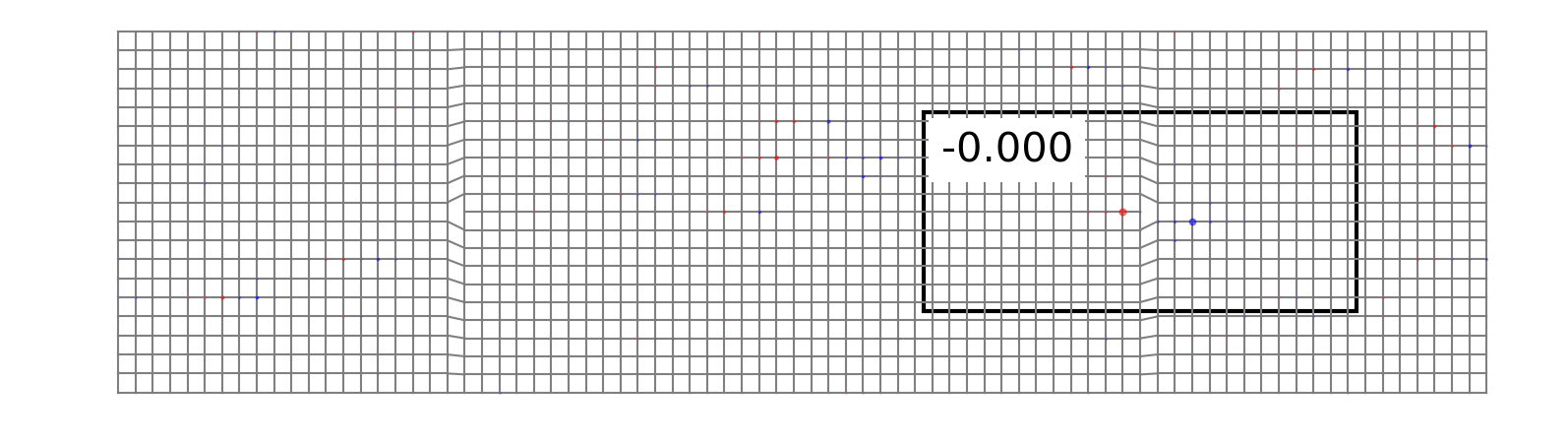}
        \label{strongDisorder1}
    }
    
    \subfloat[$\hat H_2$ at the point $(W_\lambda,W_\gamma)=(2,2)$, where $\nu_x=1$.]{
        \includegraphics[width=.9\columnwidth]{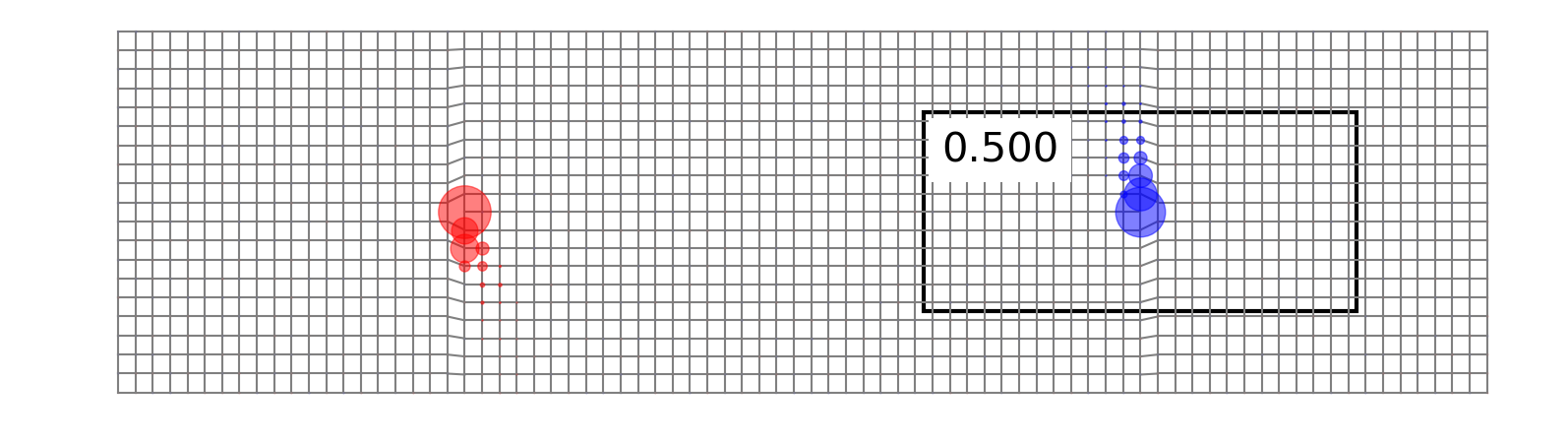}
    }
    
    \subfloat[$\hat H_2$ at the point $(W_\lambda,W_\gamma)=(22,2)$, where $\nu_x=1$.]{
        \includegraphics[width=.9\columnwidth]{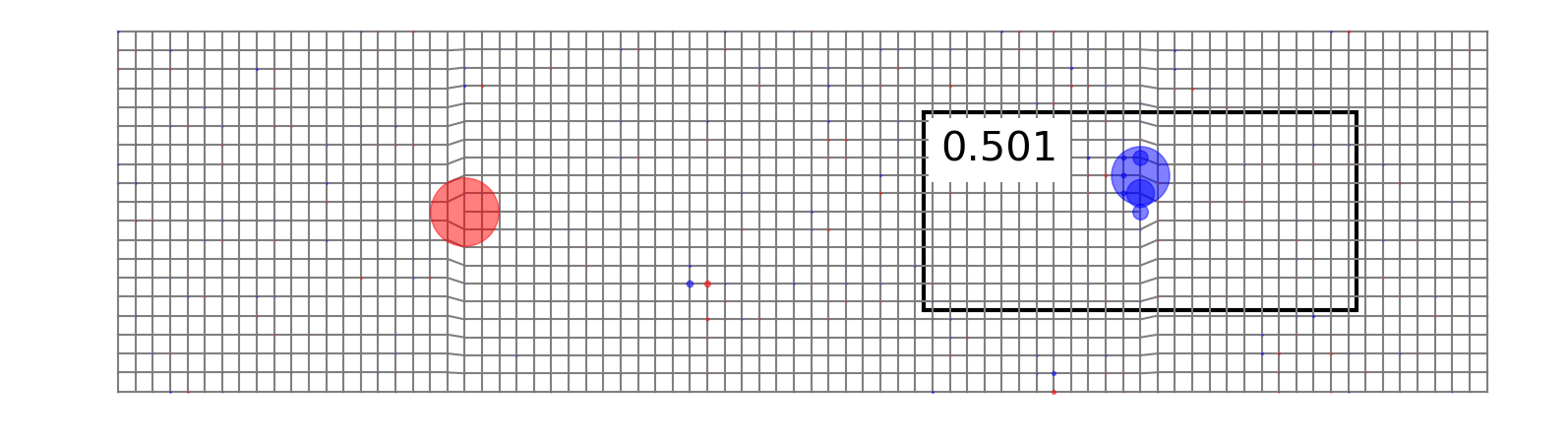}
        \label{strongDisorder2}
    }
    \caption{Excess or deficit charge densities near dislocations, for models (a,b) $\hat{H}_1$ and (c,d) $\hat{H}_2$ at the points in the phase diagram marked by squares in Figs. \ref{resultsModel1}a and \ref{resultsModel2}a respectively. In all cases, we see $\nu_x$ determines the presence or absence of bound charge at the dislocation site.}
    \label{dislocations}
\end{figure}

We confirm this behavior persists in the presence of bulk disorder by calculating the charge density near an edge dislocation, as shown in Fig. \ref{dislocations}. There, we plot the charge density near an edge dislocation for the points in the phase diagram indicated by black squares in Figs. \ref{resultsModel1}a and \ref{resultsModel2}a. We see that for both models there is a charge of $\pm 1/2$ bound to the defect in the $\nu_x=1$ phase, and no charge in the $\nu_x=0$ phase. Remarkably this is true even for the very strong disorder case shown in Fig. \ref{strongDisorder2} where the system has $\nu_x=1$ for large values of disorder that are not perturbations around a clean limit. We conjecture that this behavior is generic, i.e., Eq. \ref{boundChargeEqn} remains valid for systems with arbitrary disorder, provided we are in a region of quantized $\nu_\alpha$ where the localization length at $E_F=0$ is finite. We remark that in the region of delocalized states it was not possible to obtain reliable calculations of the localized dislocation charge, so the behavior of dislocations in this part of the phase diagram remains an open question.

\section{Discussion/Conclusion}

By extending real-space formulas for the strong topological invariant, we have introduced a method to define real-space weak topological invariants for AIII systems. Our invariants are quantized at weak disorder, as in Refs. \onlinecite{prodan2016generalized,bourne2018application}. By combining our real-space formula with a numerical transfer matrix method, we have also gone beyond those works and demonstrated that for the AIII class in regions without delocalized states at $E=0$, the weak indices take on quantized values. In contrast to the 1D case, the weak topological indices do not transition at a critical value of disorder, but instead have a critical region where the topological index $\nu_x$ smoothly transitions between quantized plateaus. We conjecture that this behavior is generic for homogeneous disorder, although there are no known analytic proofs of the quantization of weak invariants in general. The critical region takes the place of the intermediate 2D Dirac semi-metal phase that separates trivial insulators from weak topological insulators in clean systems. Indeed, our results share some similarities with calculations done for disordered 3D Weyl semi-metals that are proximate to a weak topological insulator phase\cite{shapourian2016}. In addition, we have connected the weak invariants to real-space observables. Like the clean system, we see that nontrivial disordered systems bind a quantized amount of charge at dislocation cores at both weak and strong disorder. Such systems also display a quantized polarization. We conjecture that this behavior is also generic.

One could extend this method to weak topological phases in other symmetry classes and dimensions where the formula for the corresponding real-space strong topological index is known. In general, a combination of real-space calculations of the topological index and computation of the localization length at the Fermi level should be able to precisely map the phase diagram in disorder-space in many cases. It would also be interesting to see if the link between these weak topological indices and bound states at dislocations generalizes to other symmetry classes, for example the existence of chiral/helical states bound to 2D line defects in the 3D A/AII classes\cite{ran2009one}.

\section*{Acknowledgements}
TLH thanks the National Science Foundation (NSF) grant DMR-1351895
(CAREER) for support.
TLH also thanks the National Science Foundation under Grant No. NSF PHY-1748958(KITP) for partial support at the end stage of this work during the Topological Quantum Matter program. This work made use of the Illinois Campus Cluster, a computing resource that is operated by the Illinois Campus Cluster Program (ICCP) in conjunction with the National Center for Supercomputing Applications (NCSA) and which is supported by funds from the University of Illinois at Urbana-Champaign.

\bibliography{bibliography.bib}

\end{document}